\documentclass[aip,graphicx,floatfix,preprint,superscriptaddress]{revtex4-1}

\usepackage{amsfonts}
\usepackage{amsmath}
\usepackage{graphicx}
\usepackage{longtable}

\draft 

\begin{document}


\title{Development and application of a particle-particle particle-mesh Ewald
method for dispersion interactions} 



\author{Rolf E. Isele-Holder}
\email[]{isele@aices.rwth-aachen.de}
\affiliation{Aachener Verfahrenstechnik: Molecular Simulations and Transformations,
Faculty of Mechanical Engineering, and AICES Graduate School, RWTH Aachen University, Schinkelstr. 2,
52062 Aachen, Germany}

\author{Wayne Mitchell}
\affiliation{Aachener Verfahrenstechnik: Molecular Simulations and Transformations,
Faculty of Mechanical Engineering, and AICES Graduate School, RWTH Aachen University, Schinkelstr. 2,
52062 Aachen, Germany}
\affiliation{Loyola University, 6363 Saint Charles Ave, New Orleans, LA 70118}

\author{Ahmed E. Ismail}
\email[]{aei@alum.mit.edu}
\affiliation{Aachener Verfahrenstechnik: Molecular Simulations and Transformations,
Faculty of Mechanical Engineering, and AICES Graduate School, RWTH Aachen University, Schinkelstr. 2,
52062 Aachen, Germany}


\date{\today}

\begin{abstract}
For inhomogeneous systems with interfaces, the inclusion of long-range
dispersion interactions is necessary to achieve consistency between molecular
simulation calculations and experimental results. 
For accurate and efficient incorporation of these contributions, we
have implemented a particle-particle particle-mesh (PPPM)
Ewald solver for dispersion ($r^{-6}$) interactions into the LAMMPS
molecular dynamics package.
We demonstrate that the solver's $\mathcal{O}(N\log N)$ scaling behavior allows its
application to large-scale simulations.
We carefully determine a set of parameters for the solver that
provides accurate results and efficient computation. We perform
a series of simulations with Lennard-Jones particles, SPC/E water, and hexane
to show that with our choice of parameters the dependence of physical results
on the chosen cutoff radius is removed. Physical results and computation time
of these simulations are compared to results obtained using either a plain
cutoff or a traditional Ewald sum for dispersion.
\end{abstract}


\maketitle 


\section{Introduction}
Despite their weak $r^{-6}$ scaling, dispersion forces 
``play a role in a host
of important phenomena such as adhesion; surface tension; physical adsorption; wetting; 
the properties of gases, liquids, and thin films; the strength of solids; the flocculation of particles 
in liquids; and the structures of condensed macromolecules such as proteins and
polymers.''\cite{Israelachvili.2011} 
Unsurprisingly, their contributions to intermolecular interactions are accounted
for in the vast majority of the nonbonded potentials applied in molecular simulations.
Typically, dispersion interactions are only considered within a cutoff of around
1\,nm. For homogeneous systems, the contributions of long-ranged interactions can be estimated 
efficiently and accurately.\cite{CompSimLiq}
For inhomogeneous systems, however, these corrections are inaccurate, and
computational requirements has precluded the inclusion of long-range dispersion interactions,
even though, as can be seen from the applications above, 
they are especially relevant for these systems. The necessity
of incorporating the long-range effects of dispersion forces has 
been shown in numerous studies on surface simulations\cite{Chapela.1977,
Blokhuis.1995,  Guo.1997_1, Guo.1997_2, Mecke.1997,  Janecek.2006, Lopez-Lemus.2006,
intVeld.2007, Ismail.2007} of which only some are referenced here, but also in simulations
near the critical point,\cite{Ou-Yang.2005} and in simulations of protein-ligand binding.\cite{Shirts.2007}

Various correction methods have been proposed for incorporating long-range
dispersion contributions. Most molecular simulation packages already include
energy and pressure corrections assuming homogeneous systems.
Similar ``on-line'' methods that can be applied
during simulations have been presented by Guo et al.\cite{Guo.1997_1, Guo.1997_2} for Monte Carlo
and by Mecke et al.\cite{Mecke.1997} and Jane\v{c}ek\cite{Janecek.2006} for
molecular dynamics (MD) simulations.
Chapela et al.\cite{Chapela.1977} and Blokhuis et al.
\cite{Blokhuis.1995} have developed a tail correction for simulated surface tensions that can be added 
after the end of the simulations.  
The aformentioned correction methods are
applicable only to simulations with planar interfaces. The use of a twin-range
cutoff has been proposed by Lag\"ue et al.\cite{Lague.2003} Wu and
Brooks\cite{Wu.2005} present the isotropic periodic sum for electrostatic interactions,
but the method can be extended to incorporate long-range dispersion forces.

An alternative to these ``correction''-based schemes is to include the long-range interactions explicitly
using Ewald summation, which was originally developed for the treatment of
Coulomb forces.\cite{Ewald.1921}
This method was developed for dispersion by Williams,\cite{Williams.1970}
Perram,\cite{Perram.1988} and Karasawa and Goddard,\cite{Karasawa.1989}
and later applied to surface simulations
by L\'opez-Lemus et al.,\cite{Lopez-Lemus.2006} Grest and
co-workers,\cite{intVeld.2007, Ismail.2007} Ou-Yang et al.,\cite{Ou-Yang.2005}
and Alejandre and Chapela.\cite{Alejandre.2010} Among the previously mentioned
methods for treating long-range dispersion interactions, the Ewald sum is usually
the most accurate, most reliable, and most versatile. However, its
$\mathcal{O}(N^{3/2})$ scaling\cite{Perram.1988} prohibits its application in
large-scale simulations.
This problem can be overcome by applying grid-based Ewald summation methods,\cite{Hockney.1988, 
Darden.1993, Essmann.1995, Deserno.1998.1}
whose scaling, because of the use of fast Fourier transforms (FFT), is 
$\mathcal{O}(N\log N)$. 
While frequently used for Coulomb interactions, such methods have also been applied to dispersion
interaction by Essman et al.\cite{Essmann.1995} and Shi et al.\cite{Shi.2006}
In the former work, the dispersion part is only adressed very briefly
for particle-mesh Ewald (PME), and not for PPPM. The latter puts a stronger focus
on the PPPM and dispersion interactions. We feel, however, that the description is incomplete;
for example, the equations for
the energy and virial and the exact formulation of the true reference force
are not given. Furthermore, we provide a simpler method
for calculating the pressure tensor required
for calculating surface tensions, and outline reasons why
their simulated surface tensions do not agree with other reported results 
for SPC/E water.

We present the results of the development and implementation of a
particle-particle particle-mesh (PPPM) solver for $r^{-6}$ dispersion interaction in the LAMMPS\cite{LAMMPS}
molecular dynamics package. The theory is given in Section \ref{equations}.
Error estimates for the real and reciprocal space contributions,
as well as a discussion on the limits of the error estimate, 
are presented in section \ref{error_measure}. 
The scaling behavior of the developed algorithm is presented in
Section \ref{sec_scaling}. We have performed a variety of interfacial
simulations, as long-range dispersion interactions are known to have
a significant effect on simulated surface tensions. The theory for
calculating surface tensions is briefly reviewed in Section
\ref{surftens}. A set of parameters for performing successful
simulations with the PPPM for dispersion is determined in Section
\ref{influence}. We use these parameters in Section \ref{application}
to study the surface tension of Lennard-Jones (LJ) particles, hexane,
and SPC/E water. Section \ref{performance} contains a brief comparison
of the simulation time of different solvers.
We summarize our findings in the final section.

\section{Formulation of the mesh-based dispersion sum}
\label{equations}
Excellent reviews on traditional and mesh-based Ewald sums are given by
Hockney and Eastwood,\cite{Hockney.1988} Essmann et al.,\cite{Essmann.1995}
and Deserno and Holm.\cite{Deserno.1998.1}
Karasawa et al.\cite{Karasawa.1989} provide a comprehensive description of the
traditional Ewald sum for dispersion interactions.
To make the presentation as self-contained as possible, we provide a
complete summary of the PPPM algorithm as applied to dispersion interactions, compiled from
the above references.

\begin{figure}
\includegraphics[scale=1]{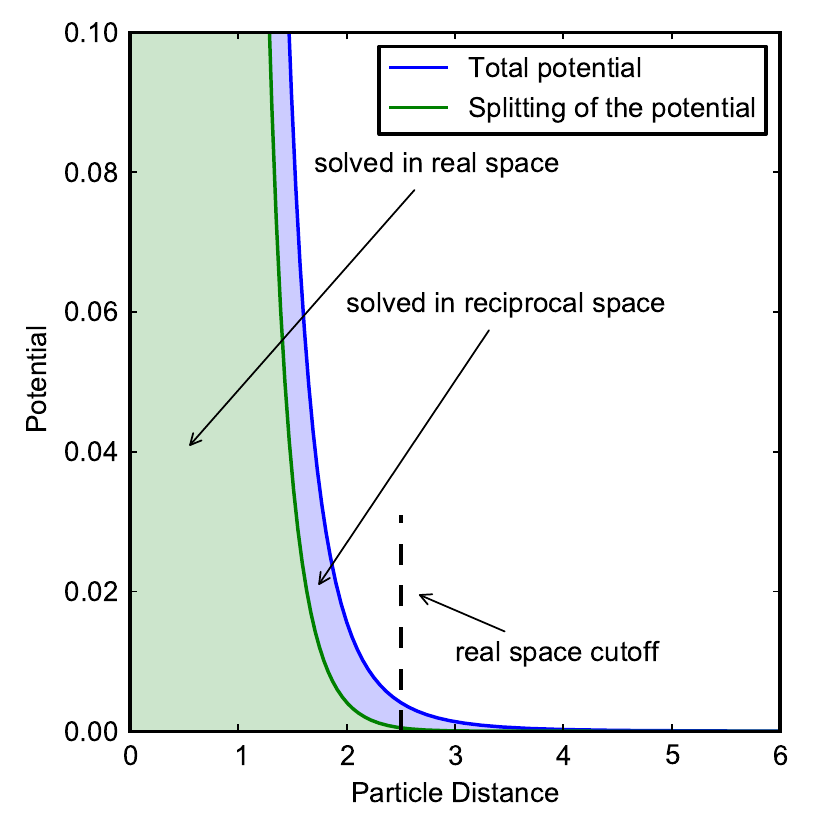}
\caption{Total potential and split potential for $r^{-6}$. When not using
the Ewald sum, the whole area under the blue curve is solved in
real space, whereas only the area under the green curve is solved
in real space when using the Ewald sum.
The error in the calculation is related to the area under
the curves beyond the cutoff. Using the Ewald technique is
thus more accurate.
\label{idea}}
\end{figure}

Because of its physical origin in the overlap of electron hulls, the repulsive
(often $r^{-n}$, where typically $n \ge 9$) part of pair potentials is very
short-ranged and can be neglected beyond a typical cutoff length of 1\,nm. We
therefore exclude the repulsive term from further consideration.
The attractive part between two atomic sites of some commonly used pair
potentials, such as the LJ or Buckingham
potential,\cite{Stone.1997} can be
expressed as
\begin{equation}
u_{\mathrm{attr}}(r_{ij}) = -\frac{C_{ij}}{r^6},
\end{equation}
where $r$ is the distance between particles $i$ and $j$ and $C_{ij}$ is the dispersion
coefficient describing the strength of the interaction.
The goal of the Ewald summation is to split this potential into a fast-decaying
potential, whose contribution can be neglected beyond a cutoff, and a
slowly decaying potential, whose contribution can be accounted for in Fourier
space, as shown in Figure \ref{idea}. Its calculation requires a Fourier transform of the dispersion
coefficients into the reciprocal space.

The benefit of mesh-based Ewald methods, such as PPPM, is that the
dispersion coefficients are distributed onto a grid, which permits application
of the FFT for the calculation of the dispersion
coefficient density. The calculation of the mesh Ewald sums in reciprocal space requires
additional steps. The dispersion coefficients have to be distributed onto
a grid and transformed into reciprocal space to calculate their interactions
in Fourier space. The resulting potential must then be derived
and backinterpolated onto the atomic sites to obtain the forces.

The dispersion energy of a system with the potential above is given
by\cite{Karasawa.1989, Deserno.1998.1}
\begin{align} \label{energy}
E_6  =  & \frac{\beta^6}{2} \sum_{i,j} C_{ij}\left( 1 + \beta^2
r_{ij}^2 + \frac{\beta^4 r_{ij}^4}{2}\right)
\frac{\exp\left(-r_{ij}^2\beta^2\right)}{\beta^6 r_{ij}^6}   \nonumber  \\
        & + \frac{1}{2V} \sum_{\mathbf{k} \in \mathbb{M}}
        \hat{G}(\mathbf{k}) \hat{\mathcal{S}}_6^2(\mathbf{k})  \nonumber\\
        & + \frac{\pi^{3/2} \beta^3}{6V}\sum_{i,j} C_{ij} - \frac{\beta^{6}}{12}\sum_{i} C_{ii},
\end{align}
where $\beta$ is the Ewald parameter for dispersion interactions,
$r_{ij}$ is the distance between particle $i$ and the nearest image of particle $j$,
and $V$ is the volume of the simulation box. $\hat{G}$ is the optimum 
dispersion influence
function, which has a different form than the electrostatic influence function.
$\hat{\mathcal{S}}_6^2$ is a function of the location and strength of the dispersion sites. 
The $\mathbf{k}$ vectors
form the discrete set $\left\lbrace 2\pi \mathbf{n}/L \right\rbrace$, where $L$
is the length of the box vectors and the components of ${\bf n} = (n_x, n_y,
n_z)$ are integer values that are zero for the center node
of the grid.
The first sum in 
equation \ref{energy} is over all pairs of atoms. However, as the potential
decays quickly with increasing interparticle distance, it
is only evaluated for particles whose $r_{ij}$ is smaller than a chosen cutoff.
The second sum is over the reciprocal vectors of all grid points.

The expression for the optimum influence function, which minimizes
the error in the calculated forces, is\cite{Hockney.1988, Deserno.1998.1}
\begin{equation}
\label{influence_function}
\hat{G}(\mathbf{k}) = \frac{\mathbf{\tilde{D}}(\mathbf{k}) \sum_{\mathbf{m} \in \mathbb{Z}^3} 
                                  \tilde{U}^2 \left( \mathbf{k} + \frac{2\pi}{h}\mathbf{m} \right)
                                  \mathbf{\tilde{R}} \left( \mathbf{k} + \frac{2\pi}{h}\mathbf{m} \right)}
                                  {|\tilde{\mathbf{D}}(\mathbf{k})|^2
                                  \left[ \sum_{\mathbf{m} \in \mathbb{Z}^3} 
                                  \tilde{U}^2 \left( \mathbf{k} + \frac{2\pi}{h}\mathbf{m} \right)\right]^2},
\end{equation}
with 
\begin{equation}
  \tilde{U}(\mathbf{k}) = \frac{\tilde{W}^{(P)}(\mathbf{k})}{V},
\end{equation}
where $P$ is the interpolation order and
\begin{equation}
 \tilde{W}^{(P)}  = h^3\left( \frac{\sin (k_xh/2)}{k_xh/2}
                          \frac{\sin (k_yh/2)}{k_yh/2} 
                         \frac{\sin (k_zh/2)}{k_zh/2} \right)^P
\end{equation}
is the Fourier transform of the interpolation function $W^{(P)}$, which is 
described, for example, in Ref.\ \onlinecite{Deserno.1998.1},
and $h$ is the grid spacing. $\mathbf{\tilde{D}} $ is the Fourier
transform of the differentiation operator required for the force calculation. In this study, we use differentiation in
Fourier space
\begin{equation}
\mathbf{\tilde{D}}(\mathbf{k}) = i\mathbf{k}.
\end{equation}
$\mathbf{\tilde{R}}$ is the Fourier transform of the true reference force and 
can, for dispersion interaction, be calculated as\cite{Deserno.1998.1,
  Essmann.1995}
\begin{equation}
\mathbf{\tilde{R}}(\mathbf{k}) = i\mathbf{k}\frac{\pi^{3/2}\beta^3}{3}
                                 \left[ \left( 1-2b^2\right)e^{-b^2}                                  
                                 + 2b^3\sqrt{\pi} \mathrm{erfc}\left( b \right)
                                  \right],
\end{equation}
with $b = {|\mathbf{k}|}/{2\beta}$.

For our choice of $\tilde{U}$, the denominator in equation \ref{influence_function} 
can be evaluated analytically, as shown by Hockney and Eastwood\cite{Hockney.1988}
or in a more explicit form by Pollock et al.\cite{Pollock.1996} The sum in the numerator  
is usually sufficiently converged when terms with $|\mathbf{m}| \leq 2$ are
included.
As the influence function is independent of the particle coordinates, it
needs to be calculated only at the beginning of a simulation or when the volume
has changed.

When the dispersion coefficient of a pair of atoms can be expressed using a
geometric mixing rule,
\begin{equation}
\label{geometric}
C_{ij} = \sqrt{C_{ii}C_{jj}} = c_{i}c_{j},
\end{equation}
as in, for example, the OPLS potential,\cite{OPLS} the function
$ \hat{\mathcal{S}}_6^2(\mathbf{k}) $ can be expressed as
\begin{equation}
\hat{\mathcal{S}}_6^2(\mathbf{k}) = \hat{S}_6(\mathbf{k})\hat{S}_6^*(\mathbf{k}),
\end{equation}
where $\hat{S}_6^{*}$ is the complex conjugate of the structure factor $\hat{S}_6$,
which is the discrete Fourier transform of the dispersion coefficient density $c_M$
on the grid points:
\begin{equation}
\hat{S}_6 (\mathbf{k}) = \sum_{\mathbf{r_p} \in \mathbb{M}} c_M (\mathbf{r_p})
                             \exp \left( -i\mathbf{k\cdot r_p}\right).
\end{equation}
When using an LJ potential, the dispersion coefficients of unlike sites are often
determined via the Lorentz-Berthelot mixing rule as
\begin{equation}
\label{arithmetic}
C_{ij} = 4\sqrt{\epsilon_i\epsilon_j}\left(\frac{\sigma_i + \sigma_j}{2} \right)^6,
\end{equation}
where $\epsilon$ and $\sigma$ are LJ parameters.
Equation \ref{geometric} cannot be used in this
case; instead, $ \hat{\mathcal{S}}_6^2(\mathbf{k}) $ must be calculated
by\cite{intVeld.2007}
\begin{equation}
\label{sum}
\hat{\mathcal{S}}_6^2(\mathbf{k}) = \sum_{i=0}^6\hat{S}_{6,k}(\mathbf{k})\hat{S}_{6,k}^*(\mathbf{k}),
\end{equation}
with
\begin{equation}
\hat{S}_{6,k} (\mathbf{k}) = \sum_{\mathbf{r_p} \in \mathbb{M}} c_{k,M} (\mathbf{r_p})
                             \exp \left( -i\mathbf{k\cdot r_p}\right),
\end{equation}
where $c_{k,M}$ is the dispersion coefficient density on the mesh points obtained from 
interpolating the dispersion coefficients
\begin{equation}
c_{i,k} = \frac{1}{4}\sigma_i^k \sqrt{\binom{6}{k}\epsilon_i}
\end{equation}
onto a grid. Because of their symmetry only four of the seven addends in
equation \ref{sum} have to be calculated. Although the imaginary part of each addend is not
necessarily zero, the imaginary parts of the sum will cancel out identically.

Additional steps are required for calculating the mesh-based forces.
For $i{\mathbf k}$ differentiation, the dispersion field can be calculated
as\cite{Deserno.1998.1}
\begin{equation}
\label{dfield}
\mathbf{E}(\mathbf{r_p}) = -\overleftarrow{\mathrm{FFT}}\left(i\mathbf{k}\frac{1}{V}\hat{S}_6\hat{G}\right)(\mathbf{r_p}),
\end{equation}
where $\overleftarrow{\mathrm{FFT}}$ indicates the reverse fast Fourier
transform.
The total force on particle $i$, based on the the real and the reciprocal part,
can then be calculated as\cite{Karasawa.1989, Deserno.1998.1}
\begin{align}
\label{force}
\mathbf{F}_i = & \sum_{j} C_{ij} \left( \frac{6}{r_{ij}^8} +
\frac{6\beta^2}{r_{ij}^6} + \frac{3\beta^4}{r_{ij}^4} +
\frac{\beta^6}{r_{ij}^2} \right) \times \nonumber \\
               &  \exp\left(-r_{ij}^2\beta^2\right) \mathbf{r}_{ij} \nonumber + 
                              c_{i} \sum_{\mathbf{r_p} \in \mathbb{M}} \mathbf{E}(\mathbf{r_p}) W(\mathbf{r}_i - \mathbf{r_p}).\\
\end{align}
It should be noted that equation \ref{dfield} and the second term in equation
\ref{force} have to be calculated for each of the seven structure factors when
the Lorentz-Berthelot mixing rule is used.

The instantaneous stress is given by\cite{Karasawa.1989}
\begin{align}
V\Pi_{\alpha\beta} = & \frac{1}{2}\sum_{i,j} C_{ij}
                                  \left( \frac{6}{r_{ij}^8} +
                                  \frac{6\beta^2}{r_{ij}^6} +
                                  \frac{3\beta^4}{r_{ij}^4} +
                                  \frac{\beta^6}{r_{ij}^2} \right)
                                \times 
                                  \nonumber \\
                     &   
                                \exp\left(-r_{ij}^2\beta^2\right)
				\mathbf{r}_{ij,\alpha}\mathbf{r}_{ij,\beta} 
                                + \frac{1}{2V} \sum_{\mathbf{k} \in
                                \mathbb{M}} \hat{G}(\mathbf{k}) \hat{\mathcal{S}}_6^2(\mathbf{k}) \times \nonumber \\
                     & \left(\delta_{\alpha\beta} -
                     \frac{3}{k^2}\frac{2b^3\sqrt{\pi}\mathrm{ erfc}(b) -
                     2b^2e^{-b^2}} {2b^3\sqrt{\pi}\mathrm{ erfc}(b) + (1-
                     2b^2)e^{-b^2}}k_{\alpha}k_{\beta} \right)  \nonumber\\
                     & + \frac{\pi^{2/3}\beta^3}{6V}\sum_{i,j}C_{ij}\delta_{\alpha\beta},
\end{align}
where $\delta_{\alpha\beta}$ is the Kronecker delta.

\section{Formulation of an Error Measure}
\label{error_measure}
Several parameters can be tuned to influence the accuracy of the dispersion PPPM:
the chosen cutoff radius $r_c$ for the sum in real space, the Ewald parameter
$\beta$, the grid size, and the order of the interpolation function for distributing
the dispersion coefficient onto a grid. The qualitative influence of the parameters
can be understood easily. The real space error arises from truncating the
pair potential. Increasing the cutoff radius or the Ewald parameter, which leads 
to a faster decaying real space potential, increases the accuracy in real space.
The precision in reciprocal space
depends on the Ewald parameter, the grid spacing, and the interpolation order.
Decreasing either of the first two or increasing the latter of these parameters
will lead to higher accuracy in the reciprocal space contribution.

To choose the tunable parameters effectively, a more quantitative 
understanding of the parameters' influence is required. The following sections
present estimates for the error of real and reciprocal space contribution
to the forces.

\subsection{Error measure for the real-space contribution}
To describe the real space error, we extend the derivation of Kolafa and
Perram\cite{Kolafa.1992} for Coulomb interaction to $r^{-6}$ potentials. The sum
of the square of the real-space contribution of the dispersion interaction of the particles beyond the cutoff $r_c$ on a single particle can be expressed as
\begin{eqnarray}
\Delta F_i^2 & = & c_i^2 \sum_{j: r_{ij} > r_c} c_j^2  \left( \frac{6}{r_{ij}^8}
+ \frac{6\beta^2}{r_{ij}^6} + \frac{3\beta^4}{r_{ij}^4} +
\frac{\beta^6}{r_{ij}^2} \right)^2 \times \nonumber \\
                      &   & \exp\left(-2r_{ij}^2\beta^2\right)r_{ij}^2.
\end{eqnarray}
Assuming that the particles are randomly distributed beyond the cutoff,
the sum can be replaced by an integral to arrive at
\begin{eqnarray*}
\Delta F_i^2 &=& c_i^2 \sum_{j} c_j^2 \frac{1}{V}  \int_{r_c}^{\infty}  
                        \left( \frac{6}{r_{ij}^6} +
                        \frac{6\beta^2}{r_{ij}^4} +
                        \frac{3\beta^4}{r_{ij}^2} + \beta^6 \right)^2 \times
                        \nonumber \\
                     & &\exp\left(-2r_{ij}^2\beta^2\right)4\pi \mathrm{d}r. 
\end{eqnarray*}
Using 
\[
\int_A^{\infty} \exp (-Bx^2) f(x) \mathrm{d}x \approx \exp (-BA^2) \frac{f(A)}{2BA}, 
\]
which is valid for $ B > 0 $, and
\[
\frac{\mathrm{d} (f(x)/2Bx)}{\mathrm{d} x} \leq f(x)
\]
for $x \geq A$, we arrive at
\begin{eqnarray}
\Delta F_i^2 & =& c_i^2 \sum_{j} c_j^2 \frac{\pi \beta^{10}}{Vr_c} 
                          \left( \frac{6}{r_{c}^6\beta^6} + \frac{6}{r_{c}^4\beta^4} + 
                          \frac{3}{r_{c}^2\beta^2} + 1 \right)^2   \times \nonumber \\
                     &  &\exp\left(-2r_{c}^2\beta^2\right),  
\end{eqnarray}
which leads to the averaged error in the force
\begin{eqnarray}
\label{real_error}
\Delta F_{\mathrm{real}} & = & \sqrt{\frac{1}{N}\sum_i \Delta F_i^2 } \nonumber \\
          & = & \frac{\mathcal{C}\sqrt{\pi}\beta^5}{\sqrt{NVr_c}} 
                 \left( \frac{6}{r_{c}^6\beta^6} + \frac{6}{r_{c}^4\beta^4} + 
                        \frac{3}{r_{c}^2\beta^2} + 1 \right) \times \nonumber \\
          &  &        \exp\left(-r_{c}^2\beta^2\right), 
\end{eqnarray}
where $N$ is the number of particles and
\[
\mathcal{C} = \sum_i c_i^2.
\]

\subsection{Formulation of an estimate for the reciprocal space error}

The following sections
present an estimate for the error of reciprocal space contribution
to the forces that is
an extension to $r^{-6}$ potentials of the analytical error measure by 
Deserno and Holm\cite{Deserno.1998.2} for Coulomb interactions.

Following the reasoning from Deserno and Holm,\cite{Deserno.1998.2}
the error in the forces in reciprocal space can be expressed by
\begin{equation}
\label{kspace_error}
\Delta F_{\mathrm{reciprocal}} = \mathcal{C} \sqrt{\frac{Q}{NV}},
\end{equation}
where $Q$ can, for the optimum choice of the influence function, be calculated as
\begin{align}
\label{full_estimate}
Q  = & \frac{1}{V} \sum_{\mathbf{k} \in \mathbb{M}}
       \left\lbrace  \sum_{\mathbf{m} \in \mathbb{Z}^3} 
       \left|\mathbf{\tilde{R}} \left( \mathbf{k} + \frac{2\pi}{h}\mathbf{m}
       \right)\right|^2 \right. \nonumber \\
   & - \left. \frac{|\mathbf{\tilde{D}}(\mathbf{k}) \sum_{\mathbf{m} \in
   \mathbb{Z}^3} \tilde{U}^2 \left( \mathbf{k} + \frac{2\pi}{h}\mathbf{m} \right)
                                  \mathbf{\tilde{R}}^* \left( \mathbf{k} + \frac{2\pi}{h}\mathbf{m} \right)|^2}
                                  {|\tilde{\mathbf{D}}(\mathbf{k})|^2
                                  \left[ \sum_{\mathbf{m} \in \mathbb{Z}^3} 
                                  \tilde{U}^2 \left( \mathbf{k} +
                                  \frac{2\pi}{h}\mathbf{m} \right)\right]^2}
                                  \right\rbrace .
\end{align}
In the following, we will derive an approximation for $Q$ that can be rapidly 
calculated.
This approximation is restricted to cubic systems with the same number of mesh points $N_m$ 
in each direction and the $i\mathbf{k}$ differentiation
scheme employed in this study. It is based on the assumption that
$h\beta$ is small.

Like Deserno and Holm,\cite{Deserno.1998.2} we exploit the fast-decaying form
of $\mathbf{\tilde{R}} $ to make the approximation that it is sufficient to retain
only $|\mathbf{m}| = 0$ in the inner sums containing $\mathbf{\tilde{R}}$.
Following the same steps, we thus arrive at 
\begin{align}
\label{integral}
Q \approx & \frac{\pi\beta^6}{12}\int_{r=0}^{\infty}
      \Bigg[ \left(r^2 - \frac{r^4}{2\beta^2}\right) \exp \left(
      \frac{-r^2}{4\beta^2}\right) + \nonumber \\
  &    \frac{2r^5}{8\beta^3}\sqrt{\pi} \mathrm{erfc} \left( \frac{r}{2\beta}
    \right) \Bigg] ^2 \times \nonumber \\
  &     \sum_{m = 0}^{P-1} c_m^{(P)} \left(\frac{rh}{2}\right) ^{2(P+m)}
    \frac{2}{2(P+m) +1} \mathrm{d}r,
\end{align}
where $c_m^{(P)}$ are coefficients given in Table \ref{coefficients}.\cite{Deserno.1998.2}
This equation corresponds to Equation (32) in Ref.\ \onlinecite{Deserno.1998.2} before changing the sum to an integral.
Performing the integration leads to the final result
\begin{eqnarray} \label{reciprocal_error}
Q &\approx & \frac{\pi^{\frac{3}{2}}}{12}\beta^{11}\sum_{m=0}^{P-1} c_m^{\left(P\right)} \left(\frac{h\beta}{2}\right)^{2\left(P+m\right)} \frac{2}{2\left(P+m\right) +1} \times \nonumber \\
\nonumber \\
 & & \left\lbrace  \mathrm{T_1 + T_2 + T_3 + T_4 + T_5 + T_6}   \right\rbrace ,  
\end{eqnarray}
with
\begin{eqnarray*}
\mathrm{T_1} & = & \frac{\left[  2\left(P+m\right) +3 \right]!!}{\sqrt{2}}     ,\\
\mathrm{T_2} & = & -\frac{\left[  2\left(P+m\right) +5 \right]!!}{\sqrt{2}}    ,\\
\mathrm{T_3} & = & \frac{\left[  2\left(P+m\right) +7 \right]!!}{4\sqrt{2}}      ,\\
\mathrm{T_4} & = & 2^{P+m+3} \left[  2\left(P+m +3 \right)\right]!!
\times \mathcal{P}_4 ,\\
\mathrm{T_5} & = & -2^{P+m+3} \left[  2\left(P+m +4 \right)\right]!! \times
\mathcal{P}_5  ,\\
\mathrm{T_6} & = & \frac{2^{P+m+3}}{2\left(P+m+5\right)+1} \left[  2\left(P+m +5\right) \right]!!
\times \mathcal{P}_6   ,\\
\end{eqnarray*}
where $x!!$ is the double factorial function
\[
x!! =  x(x - 2)!!, \left(0\right)!! = \left(-1\right)!! = 1,
\]
and $\mathcal{P}_l$ is given by
\begin{equation}
  \mathcal{P}_l = 1-\sqrt{2}\sum_{i=0}^{P+m+l-1}\frac{\left(2i-1
  \right)!!}{\left(2i\right)!!2^{i+1}}.
\end{equation}

\begin{table*}
\caption{Coefficients required for the calculation of the reciprocal space error
estimate. (Reprinted from Ref.\ \onlinecite{Deserno.1998.2}.)}
\begin{ruledtabular}
\begin{tabular}{l c c c c c c c }
$P$ & $c_0^P $ &  $c_1^P $ &  $c_2^P $ &  $c_3^P $ &  $c_4^P $ &  $c_5^P $ &  $c_6^P $ \\
\hline
1 & $\frac{2}{3}$ & & & & & &\\
2 & $\frac{2}{45}$ & $\frac{8}{189} $ & & & & &\\
3 & $\frac{4}{945} $ & $ \frac{2}{225} $ & $\frac{8}{1\,485}$ & & & &\\
4 & $\frac{2}{4\,725}$ & $\frac{16}{10\,395}$ & $\frac{5\,528}{3\,869\,775}$ & $\frac{32}{42\,525}$ & & &\\
5 & $\frac{4}{93\,555}$ & $\frac{2\,764}{11\,609\,325} $ & $\frac{8}{25\,515}$ & $\frac{7\,234}{32\,531\,625} $
   & $\frac{350\,936}{3\,206\,852\,775}$ & & \\
6 & $\frac{2\,764}{638\,512\,875} $ & $\frac{16}{467\,775}$ & $ \frac{7\,234}{119\,282\,625} $
   & $\frac{1\,403\,744}{25\,196\,700\,375}$ & $\frac{1\,396\,888}{40\,521\,009\,375}$
   & $\frac{2\,485\,856}{152\,506\,344\,375} $ & \\
7 & $\frac{8}{18\,243\,225}$ & $\frac{7\,234}{1\,550\,674\,125} $ & $\frac{701\,872}{65\,511\,420\,975}$ 
   & $\frac{2\,793\,776}{225\,759\,909\,375} $ & $ \frac{1\,242\,928}{132\,172\,165\,125} $
   & $\frac{1\,890\,912\,728}{352\,985\,880\,121\,875} $ & $\frac{21\,053\,792}{8\,522\,724\,574\,375} $ \\
\end{tabular}
\end{ruledtabular}

\label{coefficients}
\end{table*}

\subsection{Numerical Tests}

We performed test runs to examine the accuracy of the real space
and reciprocal space error estimates. We placed 2000 LJ
particles randomly in a box with box length 15\,$\sigma$ in each
direction to create a bulk system. In order to test the error
estimates for surface systems, we placed 4000 LJ particles randomly in
a $30\times30\times10\sigma^3$ box and extended the length of the
shortest box edge to 30\,$\sigma$ afterwards without changing the
particle coordinates. We calculated the real and reciprocal space
forces on the particles for these configurations seperately using
different values for the Ewald parameter, the grid size, the
interpolation order, and the real space cutoff. Interpolation
orders $P = 3, \ldots, 6$ and $2^k$ mesh points, $k = 2,\ldots,6$ in each
direction were used. Real-space cutoffs of 2.0, 3.0, and 4.0\,$\sigma$ were used. 
The error in the forces is calculated as
\begin{equation}
\Delta F = \sqrt{\frac{1}{N} \sum_{i=1}^N (\mathbf{F}_i^{\mathrm{PPPM}} - \mathbf{F}_i^{\mathrm{exact}})^2},
\end{equation}
where $\mathbf{F}_i^{\mathrm{PPPM}} $ is the force calculated with the PPPM and 
$ \mathbf{F}_i^{\mathrm{exact}} $ is the ``exact'' force calculated
with an Ewald summation\cite{intVeld.2007} in which we used a large
cutoff and a large number of recirpocal vectors to ensure proper
conversion.

\begin{figure}
\includegraphics[scale=1]{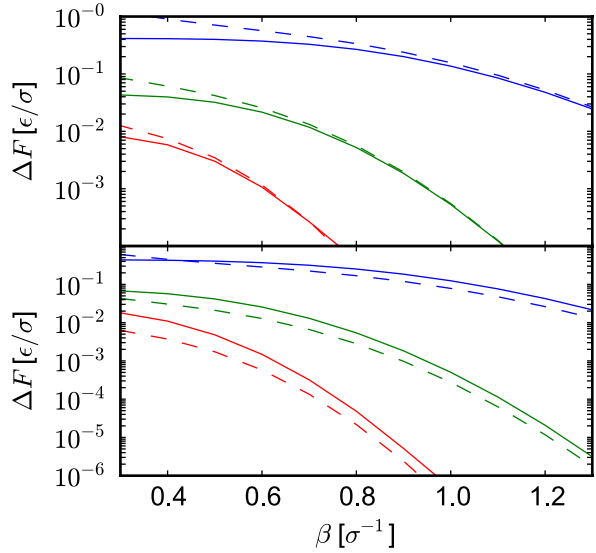}
\caption{Real space error in the forces measured and estimated with equation
  \ref{real_error} for (top) bulk system and (bottom) interfacial
  system. Measured errors are depicted as solid lines, estimated
  errors as dashed lines. From top to bottom, the real space cutoff is
  (blue) $2.0\sigma$, (green) $3.0\sigma$, and (red) $4.0\sigma$.
  The estimate works well for the bulk system, but fails
  for the interfacial system.
\label{rspace_test}}
\end{figure}

\begin{figure}
\includegraphics[scale=1]{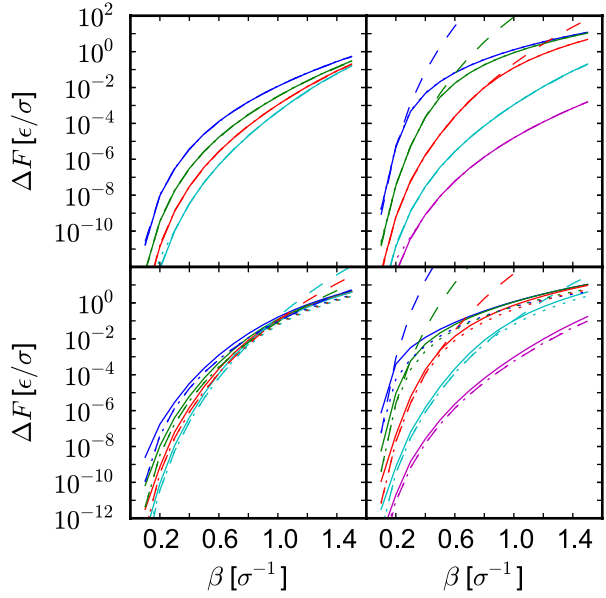}
\caption{Measured and estimated reciprocal space errors. for (top) 
  a bulk system and (bottom) an interfacial system. The graphs
  on the left show increasing
  interpolation order from $P = 3$ at the top to $P = 6$ at the bottom, with
  a fixed grid containing 32 grid points in each direction. The graphs
  on the right show grid density increasing from $2^2$ points in each direction
  at the top to $2^6$ points at the bottom, for fixed interpolation order
  $P = 5$. Measured errors are depicted as solid lines. Errors
  estimated with equations \ref{kspace_error} and \ref{reciprocal_error}
  are depicted as dashed lines, those estimated with equations
  \ref{kspace_error} and \ref{full_estimate} are depicted as dotted lines.
\label{kspace_test}}
\end{figure}

The results of the real space error estimate are given in
Figure \ref{rspace_test}. Results for the reciprocal space error estimates are
shown in Figure \ref{kspace_test}.
Except for small values of $\beta$, 
the real space error estimate
works well for the bulk system. In contrast, the error is
underestimated in surface simulations.
For bulk phase systems, the reciprocal space error estimate with
equations \ref{kspace_error} and \ref{full_estimate} provides very
good results. The approximation with equation \ref{reciprocal_error}
strongly overestimates the reciprocal space error when the assumption
that $h\beta$ is small is violated. For the interfacial system, the error
estimates underpredict the simulation error. Yet, as can be seen from
these figures and from Figure \ref{hexane_parameters}, the error
estimates can be useful for determining the value of the Ewald parameter
for which the accuracies in real and reciprocal space are equal, if this
information is needed.

The results shown above demonstrate that the error estimates
presented here should only be applied to homogeneous bulk
systems. Additionally, it should be noted that the error estimate for
the real space contribution assumes that the errors in the forces
partly cancel. This cancellation of errors cannot occur for the
real space contribution to either energy or pressure. This can be easily seen
from the following example: Consider three equal, collinear particles,
with particle 2 equidistant between particles 1 and 3.
The distance between particle 2 and the other particles
is larger than the chosen real-space cutoff, so that none of the
real-space forces, energies, or pressures are calculated.
If the chosen cutoff radius were larger, so that
the interactions should be calculated,
the force on particle 2 would be zero, because the
contributions from particles 1 and 3 cancel. The energy and the
pressure that are exerted on particle 2, however, do not cancel
but instead are additive. The reason for this behavior is that dispersion 
interactions, unlike Coulomb interactions, are always attractive.
Contributions to the energy thus always have the same sign.
As the distance vectors and force vectors for pairwise
interactions always point in opposite directions, the contributions
to the diagonal components of the virial tensor always have the same sign, too.
This in turn means that
pressures and energies can be underestimated, even when forces are
calculated accurately.

Thus, usage of the above error estimates to set the
Ewald and grid parameters is only recommended for bulk systems in
which neither the energy nor the pressure is relevant, such as in
simulations for determining diffusivities. We show how to determine
parameters for interfacial simulations in Section \ref{influence}.

\section{Scaling of the algorithm}
\label{sec_scaling}
The main benefit of mesh-based Ewald methods over traditional
Ewald sums is the improved scaling behavior of the mesh-based approach.
To examine the scalability of the implemented solver, we have performed
simulations with $2^n \times 10^3$ LJ particles, where $n=0,1,\ldots,10$,
with the dispersion PPPM solver and the Ewald summation. 
 The density was 
$3.64\sigma^{-3}$ in all simulations. The boxes were always cubic.
An energy minimization and equilibration over
50\,000 timesteps in the $NVT$ ensemble at a reduced temperature 
$T^* = 0.85$ 
 was followed by a simulation over 1\,000 timesteps
in the $NVE$ ensemble. The simulation time of the last 1\,000 timesteps 
was used to measure the performance. These simulations were executed on a single
core of an Intel Harpertown E5454 processor with eight 3.0 GHz Xeon cores.

Automated parameter generation was applied in simulations with the
Ewald sum.\cite{intVeld.2007} The mesh parameters for the PPPM were
set using the error estimate presented in the previous section in the following way.
The real space cutoff was chosen as 3.0\,$\sigma$. The real space error estimate
was then used to set the Ewald parameter to obtain a desired accuracy of 
0.01\,$\epsilon/\sigma$ in the calculated real space forces. The interpolation order was set to $P = 5$.
Using these data, the grid spacing was chosen in a way that the accuracy of
the reciprocal space forces was smaller then 0.01\,$\epsilon/\sigma$ by using 
equation \ref{reciprocal_error}. As the conditions for the validity of the 
error estimates are fulfilled for the chosen simulations, the comparison we draw
here is for different system sizes run with the same accuracy.

\begin{figure}
\includegraphics[scale=1]{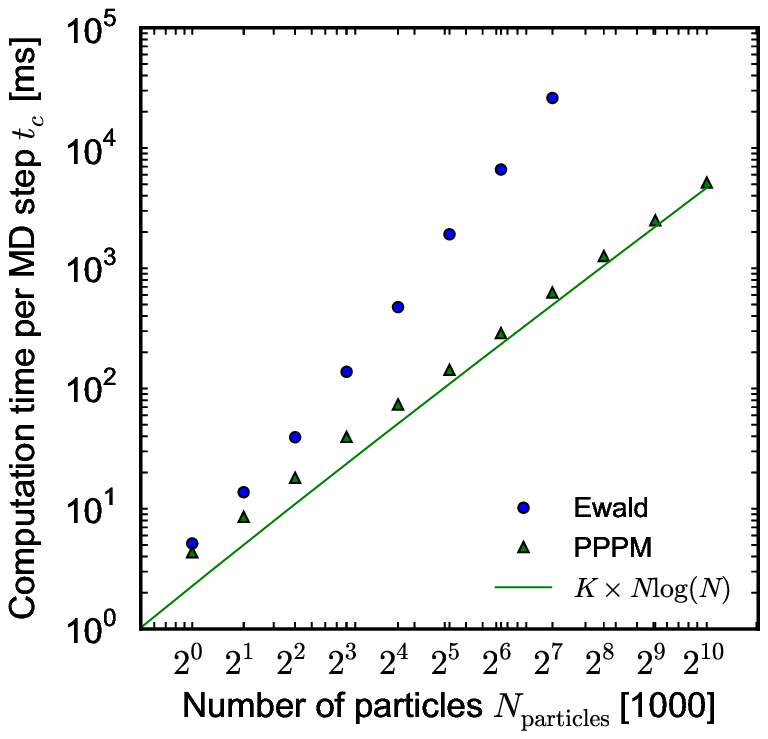}
\caption{Scaling of the Ewald sum and the PPPM method for dispersion interactions\label{scaling}}
\end{figure}

As can be seen from Figure \ref{scaling}, which shows the computation
time per timestep, the dispersion PPPM approaches
the expected scaling behavior of $\mathcal{O}(N\log N)$ with increasing
numbers of particles. Its performance becomes several magnitudes faster 
than the traditional Ewald sum
and is thus far more suitable for large-scale simulations. The comparison between
the different solvers drawn here should be considered qualitative, as we did
not examine whether the two different solvers were run with the same 
accuracy.

\section{Determination of surface tensions from MD simulation}
\label{surftens}

As the need for incorporating long-range dispersion is especially acute for interfacial
systems, we have run simulations with explicit interfaces on LJ particles, SPC/E
water, and hexane to test the efficiency and accuracy of the
dispersion PPPM algorithm. This section briefly summarizes the method applied to simulate surface tensions. 
Surface tensions can be obtained from MD simulations via two-phase
simulations. We use the approach, developed by Tolman\cite{Tolman.1948}
and Kirkwood and Buff,\cite{Kirkwood.1949} in which the surface tension is
expressed via 
\begin{equation}
\label{Kirkwood}
\gamma_p = \frac{1}{2}\int_{-\infty}^{\infty} 
                     \left( p_{\perp}(z) - p_{\parallel}(z) \right) \mathrm{d}z,
\end{equation}
where $p_{\perp}(z) = p_z(z)$ is the pressure component
normal to the surface and 
$p_{\parallel}(z) = (p_x(z) + p_y(z))/2$ is the pressure component parallel to the surface.
Replacing the integral with an ensemble average leads to
\begin{equation}
\label{ensemble}
\gamma_p = \frac{L_z}{2}\left(p_{\perp} - p_{\parallel} \right) =
                 \frac{L_z}{2}\left[ \left\langle p_z \right\rangle - 
                 \frac{\left\langle p_x \right\rangle + \left\langle p_y \right\rangle}{2} \right],
\end{equation}
where $L_z$ is the box dimension in the $z$-direction. The outer factor of
$1/2$ takes into account that the simulated system contains two interfaces.

If a cutoff is introduced for the pair potential, the surface tensions calculated with Equation
\ref{ensemble} will underestimate the correct surface tension of the simulated
material. This error can be estimated by adding a ``tail correction'' 
$\gamma_{\mathrm{tail}}$ to the simulated surface tension to provide a better
estimate of the correct surface tension
\begin{equation}
\gamma \approx \gamma_p + \gamma_{\mathrm{tail}}
\end{equation}
from the simulation. The correction can be calculated as\cite{Chapela.1977,
Blokhuis.1995}
\begin{eqnarray}
\gamma_{\mathrm{tail}}  &= &
 \frac{\pi}{2} \int_{-\infty}^{\infty} \int_{-1}^{1} \int_{r_c}^{\infty}
 r^3 \frac{\mathrm{d}U(r)}{\mathrm{d}r}g(r)(1-3s^2) \nonumber \\
 &  &  \times
 \left( \rho (z) \rho(z - sr) - \left( \rho_{\mathrm{G}}(z) \right)^2 \right)
 \mathrm{d}r \mathrm{d}s \mathrm{d}z,
\end{eqnarray}
where $U(r)$ is the pair potential, $g(r)$ is the radial distribution
function, $\rho (z)$ is the simulated density profile, $r_c$ is the 
cutoff radius for the pair potential, and $\rho_{\mathrm{G}}(z)$ is the
Gibbs dividing surface
\begin{equation}
\rho_{\mathrm{G}}(z) = \rho_c + \frac{\Delta \rho}{2} \mathrm{sgn}(z),
\end{equation}
where $\rho_c$ is the mean and $\Delta \rho$ is the difference of the 
densities of the coexisting phases.
$g(r)$ was assumed to be unity beyond the cutoff in the calculations
of the tail correction. The values for $\rho_c$ and $\Delta \rho$,
which were also used to calculate the liquid and vapor densities 
in this study,
were obtained from fitting an error function to the
simulated density profile.\cite{Huang.1969, Beysens.1987, Sides.1999}

\section{Influence of the Ewald and grid parameters on physical properties}
\label{influence}
The parameters used by the dispersion PPPM have a strong influence on both
the efficiency and the accuracy of the simulations. 
As the presented error estimates fail to describe systems with interfaces,
we have run test simulations to determine a set of parameters that can provide
both accurate results and acceptable performance for interfacial
simulations.
These parameters were
determined for Lennard-Jones particles and hexane, a nonpolar
fluid whose intermolecular interactions are dominated by dispersion.
Hexane was modeled using the OPLS-AA\cite{OPLSAA_orig} force field.

Simulations with hexane contained 689 hexane molecules that 
were placed using Packmol\cite{Packmol} in a subvolume around the center of the box with volume 
$50\times 50\times 150$\,\AA$^3$.
After an energy minimization with a soft potential and several runs
with restricted movement of the particles, the simulations were
equilibrated for 1\,000\,000\,timesteps with a timestep of
$\Delta t = 1$\,fs. The temperature was set to $T = 300$\,K using
a Nos\'e-Hoover\cite{Const_press_alg2} thermostat with
a damping factor of 0.1\,ps. A PPPM\cite{Hockney.1988} with
a real space cutoff of $r_c = 10$\,\AA, an Ewald
parameter of $\beta = 0.17$\,\AA$^{-1}$\ and fifth-order interpolation ($P = 5$)
was used to calculate the electrostatic potential. The grid dimension
was set to $20\times20\times45$.

The parameters of the PPPM for dispersion are the real space cutoff,
the Ewald parameter, 
the interpolation order, and the grid spacing in each dimension.
The influence of the different parameters is already described
at the beginning of Section \ref{error_measure}.
Instead of exploring this six-dimensional parameter space, we 
set the interpolation order to $P=5$ and the real space cutoff
$r_c = 10.0$\,\AA\ for the hexane system. 
This choice of parameters was made because these
values are commonly used in MD simulations,
although they are in principle arbitrary. We do not claim that these
are the optimal choices. For example, using the long-range
dispersion solver allows experimenting with smaller
values for the real space cutoff and might in this way improve
the performance of the calculations.
Furthermore, the grid spacing
was equal in all three dimensions in the simulations described below,
as near cubic grids usually provide most accuracte calculations.

This reduction of the parameter space allows for
determining suitable simulation parameters with less effort,
but permits reaching a wide range of accuracy in either real or reciprocal
space. As the real space cutoff is fixed, the real space accuracy
depends only on the Ewald parameter, which is therefore used
in the following simulations to tune the real space accuracy.
In principle, we could also have fixed the Ewald parameter
beforehand and modified the real-space cutoff in our simultations
to tune the real-space accuracy, but we decided against it
to have better control over the real space calculation time.
For a given Ewald parameter and the other parameters fixed,
the grid spacing can be altered to tune the reciprocal 
space accuracy, even though the reachable reciprocal-space accuracy
is not unlimited for a fixed Ewald parameter.

We performed surface tension calculations with different settings
for the two remaining parameters, the Ewald parameter and
the uniform grid spacing. We examined the resulting surface tensions
and liquid densities. In addition we determined the rms error in the total
forces as well as the real and reciprocal space contributions to the error
by comparing
the forces calculated for a single snapshot of an equilibrated systems
to forces that were calculated using a large real-space cutoff and a very small
grid spacing.

\begin{figure}
\includegraphics[scale=1]{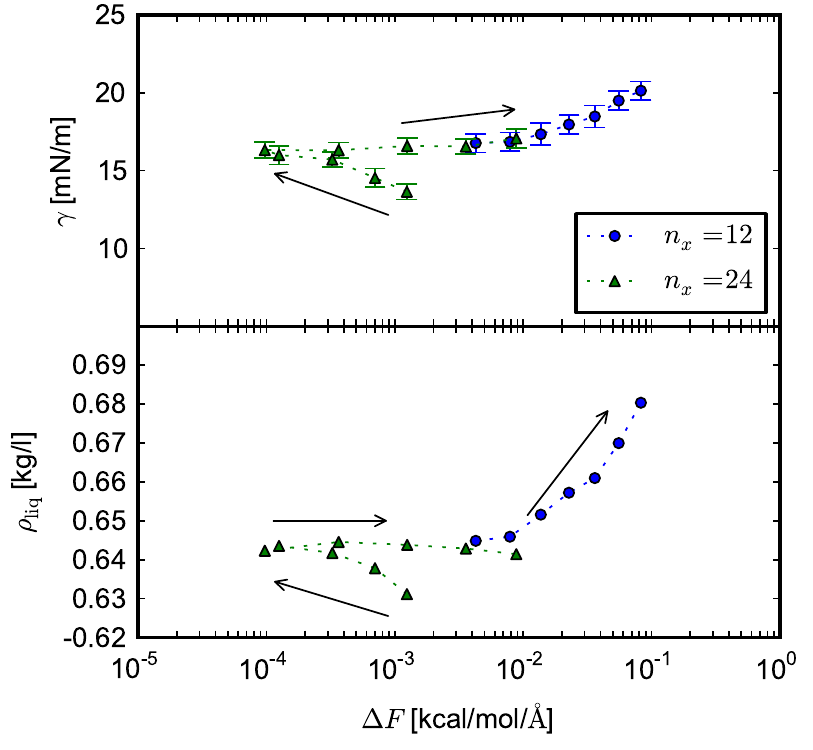}
\caption{Surface tension and density of hexane as a function of the
total error in the calculated forces. The arrows point in the direction
of increasing Ewald parameter.
\label{hexane_parameters_2}}
\end{figure}

\begin{figure}
\includegraphics[scale=1]{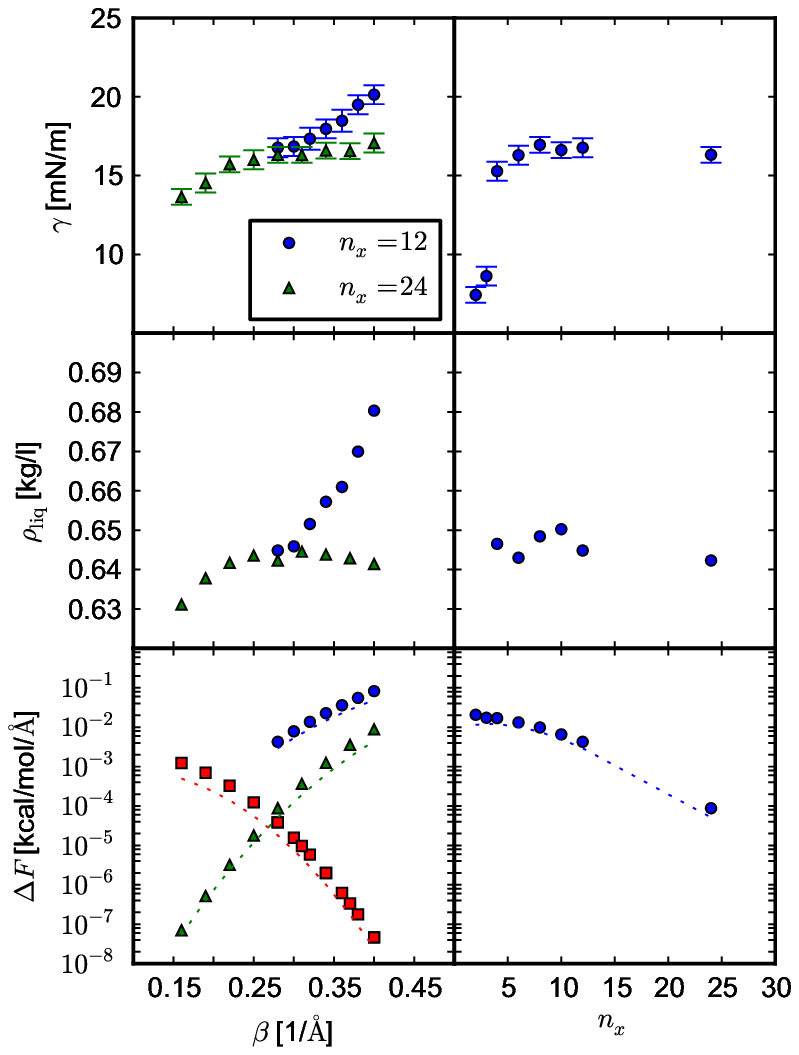}
\caption{Surface tension, density, and errors in the forces in simulations of hexane.
In the lower left graph, the red squares correspond to the real space error, while the 
triangles and circles correspond to the reciprocal space error when using the fine grid and coarse
grids, respectively. The circles in the lower right
graph correspond to the reciprocal space error. The Ewald parameter is
$\beta = 0.28$\,\AA$^{-1}$ in all figures on the right side.
Dotted lines are error estimates calculated with Equations
\ref{real_error}, \ref{kspace_error}, and \ref{full_estimate}.
\label{hexane_parameters}}
\end{figure}

The results for the surface tension and density of hexane
are given as a function of the total rms error
in the forces in Figure \ref{hexane_parameters_2}. 
In simulations with fewer grid points, the total error is always
dominated by the reciprocal space error. In simulations with
smaller grid spacings, where the number 
of grid points in the $x$-direction $n_x = 24$, the real and reciprocal space error 
are approximately equal for the highest achieved total accuracy at 
$\beta = 0.28$\,\AA$^{-1}$. As can be seen from Figure \ref{hexane_parameters}, 
the real space error dominates for smaller values of 
the Ewald parameter $\beta$, whereas
the reciprocal space error dominates for larger values of $\beta$.

As the total error decreases, the simulated surface tensions and densities plateau, 
indicating that further increases in accuracy, which can be obtained
by using even finer grids and larger values for the Ewald parameter,
will offer little benefit in the 
accuracy of the measured quantities. 
Decreasing the Ewald parameter, thereby increasing the real space error, strongly 
influences the simulated quantities. In contrast, increasing the Ewald parameter
and in this way increasing the reciprocal space error has less influence on the results.
Physical data begin to change for reciprocal space errors above approximately
0.01\,kcal mol$^{-1}$ \AA$^{-1}$. For the examined quantities, the
real space error has a stronger influence on the results than the
reciprocal space error. The reason for this observation is that an
increasing real space error leads to increasing underprediction of the
cohesion of a simulated system. For simulations of quantities in which
the cohesion does not influence the reults, the influence of the real
and reciprocal space error will possibly be different.

The data given in Figure \ref{hexane_parameters_2} are also given on the left
side of Figure \ref{hexane_parameters} as a function of the Ewald parameter.
These results, in combination with those from Figure \ref{hexane_parameters_2},
show that an Ewald parameter of approximately $\beta = 0.28$\,\AA$^{-1}$ in combination
with a real space cutoff $r_c = 10$\,\AA\ provides a sufficient real
space accuracy for the performed simulations.

As the results from Figure \ref{hexane_parameters_2} indicate that
increasing the reciprocal space error does not alter the 
obtained physical data strongly, we have performed further simulations
with fixed Ewald parameter with varying grid spacings.
Results of these simulations are given on the right side of Figure \ref{hexane_parameters}.
Increasing the number of grid points $n_x$ in the $x$-direction beyond 12 does not alter
either the simulated density or surface tensions, although the error in the forces 
continues to decrease. However, the extended running times required for the 
finer meshes make these higher fidelity calculations computationally undesirable.

Therefore, we choose $\beta = 0.28$\,\AA$^{-1}$ and the
grid spacing $h \approx 4.17$\,\AA\ as these
parameters provide sufficient accuracy. Examining the influence
of the parameters of the LJ system provided similar results
as those described above. The parameters we obtain for
the LJ system are $\beta = 1.1\sigma^{-1}$ and
$h \approx 1.22\sigma$ for Interpolation order $P=5$ and
a real space cutoff of $r_c = 3\sigma$. 
The corresponding simulations and results are described in
the supporting information \cite{support}. 

\section{Application of the solver}
\label{application}
To compare our algorithms to existing implementations---a 
plain cutoff or the Ewald sum\cite{intVeld.2007}---we have performed
simulations with systems of LJ particles, SPC/E water,\cite{SPCE}
and hexane modeled with the OPLS all-atom force field.\cite{OPLSAA_orig}
These systems cover a model system as well as  realistic systems
in which Coulomb interactions (water) and dispersion
interactions (hexane) dominate. Furthermore, these systems have already
been studied and allow comparison to results from the literature.\cite{intVeld.2007,
Ismail.2007, Chen.2007, Wang.2009, Sakamaki.2011}
A comparison with results from Shi et al.\cite{Shi.2006} is of special interest,
as they have also used a PPPM dispersion method to determine the surface tension of SPC/E water.

\subsection{Lennard-Jones particles}
\label{surftens_LJ}

The Lennard-Jones simulations were performed in a box with volume 
$11.01\times 11.01\times 176.16\sigma^3$ and 4000 particles that
were placed randomly in a subvolume at the center of the box.
After minimization using a soft
potential, the system was equilibrated for 100\,000 timesteps.
The timestep was set to 0.005\,$\tau$, where $\tau = \sigma
\sqrt{m/\epsilon}$.
Simulations were executed at reduced temperatures $T^* =
k_{\mathrm{B}}T/\epsilon \in \{ 0.7, 0.85, 1.1, 1.2 \}$
using a Nos\'e-Hoover\cite{Const_press_alg2} thermostat with
damping factor 10\,$\tau$. The equations of motion were
solved using a velocity Verlet algorithm.\cite{Verlet}
Afterwards, simulations were run for another 1\,000\,000 timesteps with
the same conditions. During that time, instantaneous surface tensions
were calculated every timestep. Configurations 
were stored every 1\,000 timesteps
to calculate the density profile.
For simulations without a long-range dispersion solver, we examined cutoffs of
$2.5\sigma$, $5\sigma$, and $7.5\sigma$.
Simulations with an Ewald solver were performed with
cutoffs of $3\sigma$, $4\sigma$, and $5\sigma$. We relied on automatic
generation of the Ewald parameter and the cutoff for the
$\mathbf{k}$ vectors. We used the value of 0.05 as
the desired relative accuracy in the forces.\cite{intVeld.2007}
The resulting Ewald parameters were $0.60\sigma^{-1}$, $0.45\sigma^{-1}$, and
$0.36\sigma^{-1}$; the number of $\mathbf{k}$ vectors were 1616, 677, and
320 for the different cutoffs.
In simulations with the dispersion PPPM we used cutoffs of $3\sigma$, $4\sigma$,
and $5\sigma$.  We used $P = 5$, $\beta = 1.1\sigma^{-1}$ and a grid with
$9\times 9\times 144$ mesh points in agreement with our results from Section \ref{influence}.

\begin{table}
\caption{Results of the validation runs for the LJ particles. Uncertainties
given in parentheses}
\scalebox{0.7}{
\begin{tabular}{l l c r r r r r}
\hline \hline
& & & & \multicolumn{3}{c}{Surface tension,
$\epsilon\sigma^{-2}$}  & \\
$T^*$  &  solver  & $r_c$ ($\sigma$)  &  $\rho_{\mathrm{liq}}$  ($\sigma{-3}$) & $\gamma_{p}  $  &  $\gamma_{t}  $ &  $\gamma $
& $t_c$ (ms)\\
\hline
0.7  & cutoff & 2.5  &    0.7865 &  0.588(30) & 0.327 &
0.915(30) & 3.7 \\
      &            & 5.0  &   0.8349   &  1.006(30) & 0.125 &
      1.131(30) & 24.8\\
      &             & 7.5  &   0.8390  & 1.112(30) &  0.057 &
      1.169(30) & 80.2\\
      & Ewald & 3.0 &   0.8332  &  1.085(30) & -        &
      1.085(30) & 123.2 \\
      &                 & 4.0 &  0.8371   & 1.121(30) & -
      & 1.121(30) & 77.5\\
      &                 & 5.0 &  0.8393  & 1.134(30) & -
      & 1.134(30) & 87.63 \\
   &  PPPM & 3.0 & 0.8404& 1.158(30) & -        &1.158(30)
   &  19.59\\
      &       & 4.0    & 0.8407 &  1.167(30)& -        & 1.167(30)
     & 32.83 \\
      &                 & 5.0   &0.8408 &1.157(30) & -
      & 1.157(30) & 52.67\\ 
      &                 &        &              &         
      &                  &          &                 \\
0.85 & cutoff & 2.5  & 0.6996   & 0.341(22)  & 0.221  &
0.562(22) & 3.1 \\
      &                & 5.0  & 0.7672   & 0.700(26)  &
      0.098  &   0.798(26) & 22.6\\
      &                & 7.5  & 0.7730  & 0.781(32)  &
      0.046  &  0.827(32) & 73.2\\
      & Ewald & 3.0  &  0.7651  & 0.742(24)  & -  &
      0.742(24) & 122.2 \\
      &                 & 4.0  &  0.7706  &  0.799(26) & -
      & 0.799(26) & 75.0 \\
      &                 & 5.0  &  0.7732  & 0.803(28)  &
      -  & 0.803(28)  & 76.0\\
      & PPPM & 3.0 & 0.7748 & 0.817(26) & -  & 0.817(26)
      & 18.81\\
      &           & 4.0    &  0.7758 & 0.829(24) & -
      & 0.829(24) & 30.00\\
      &                 & 5.0   & 0.7756 & 0.829(28) & -
      & 0.829(28) & 48.24\\ 
      &                 &        &              &             &   
      &          &    \\
1.1 &     cutoff & 2.5  & n.a.  & 0.023(26)  & n.a.  &
0.023(26) & 1.4 \\
      &                & 5.0  & 0.6282   &  0.278(26) &
      0.042  & 0.320(26) & 15.5 \\
      &                & 7.5  &  0.6385  &  0.293(24) &
      0.026  & 0.319(24) & 50.7\\
      &     Ewald & 3.0  &  0.6243  & 0.270(26)  & -  &
      0.270(26) & 118.2\\
      &                 & 4.0  &   0.6354  & 0.293(24)  & -
      & 0.293(24) & 68.4\\
      &                 & 5.0  &   0.6452  & 0.315(26)  & -
      & 0.315(26) & 56.2\\
       &     PPPM & 3.0 & 0.6451 & 0.314(24) & -  & 0.314(24)
       &  15.69\\
      &           & 4.0    & 0.6448 & 0.330(26) & - & 0.330(26)
      & 23.57 \\
      &                 & 5.0   & 0.6462 & 0.302(22) & -
      & 0.302(22) & 37.19 \\ 
      &                 &        &              &             &   
      &          &     \\
1.2 &    cutoff & 2.5  &  n.a.  &  0.001(20) & n.a.  &
0.001(20) & 1.4\\
      &                & 5.0  &  0.5613  &  0.113(20) &
      0.025  & 0.138(20) & 11.8\\
      &                & 7.5  & 0.5725  & 0.159(26)  &
      0.013  & 0.172(26) & 38.8\\
      &       Ewald & 3.0  &  0.5497  & 0.128(24)  & -  &
      0.128(24) & 118.9\\
      &                & 4.0  &  0.5647  & 0.141(26)  & -
      & 0.141(26) & 65.4\\
      &                 & 5.0  &   0.5728  & 0.159(24)  & -
      & 0.159(24) & 49.9\\
      &        PPPM & 3.0 &0.5767& 0.164(24)& -  & 0.164(24)
      & 13.73\\
      &             & 4.0    & 0.5757 & 0.155(22)& -
      & 0.155(22) & 21.10 \\
      &                 & 5.0   & 0.5766 &0.154(26) &-
      & 0.154(26) & 32.24 \\ 
\hline\hline
\end{tabular}
}
\label{LJ}
\end{table}

\begin{figure}
\includegraphics[scale=1]{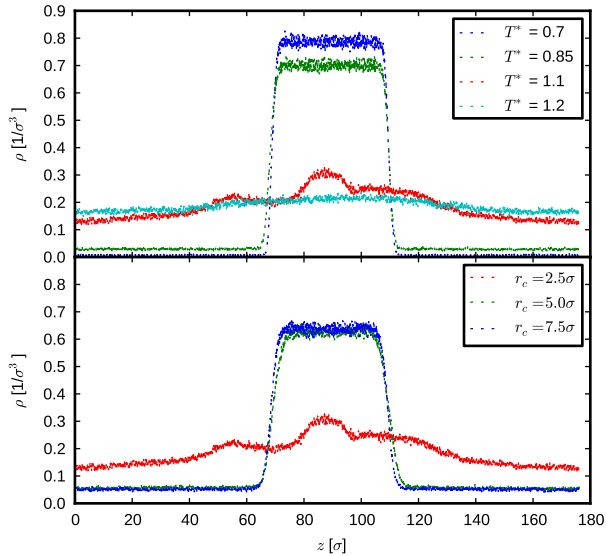}
\caption{Measured density profiles for simulations of LJ particles: 
(top) simulations with cutoff $r_c = 2.5\sigma$ and (bottom)
simulations at a reduced temperature of $T^* = 1.1$. 
\label{density_profiles}}
\end{figure}

Results are given in Table \ref{LJ}. Overall, we find good
agreement with results from the literature.\cite{intVeld.2007, Wang.2009}
The simulated densities and surface tensions
show a strong dependence on the chosen cutoff in
simulation without a long-range dispersion solver. 
For simulations at higher 
temperatures, systems with small cutoffs were so close to the critical point
that error functions were no longer appropriate for describing the density
profile, as can be seen from Figure \ref{density_profiles}.
Agreement between
simulated data with and without long-range dispersion solver can only
be obtained when using a large cutoff in simulations without the
long-range solver.

Unlike the simulations with a long-range cutoff, the results for the dispersion PPPM method do not show a dependence on the switching radius.  
For the Ewald sum, a slight dependence of the physical data remains, 
which we attribute to the automated parameter generation
routine in combination with the specified accuracy. 

\subsection{SPC/E water}
\label{spceruns}

Simulations with SPC/E water were performed with 5\,000 water molecules
in a box of $50\times 50\times 150$ \AA$^3$. 
The initial configurations of the particles were created using
Packmol.\cite{Packmol} If not explicitly given in the
following, the simulation settings were as those for hexane described in
Section \ref{influence}.

Simulations were executed at 300, 350, and 400\,K.
For each solver, we used cutoffs of 10, 12, and 16\,\AA\ for
the sum in real space for dispersion and Coulomb interaction.
The SHAKE algorithm\cite{SHAKE} was
used to constrain the bond lengths and bond angles.

A PPPM\cite{Hockney.1988} solver was used for long-range electrostatics in simulations
with a plain cutoff for dispersion interaction. We picked interpolation
order $P = 5$ and a grid of $24\times 24\times 54$ mesh points as grid
parameters.
The Ewald parameter was $\beta$ = 0.255, 0.226, and 0.184 \AA$^{-1}$ for the three
different cutoffs.
In simulations with the traditional Ewald sum for dispersion and
Coulomb interactions, 
the Ewald parameter and number of $\mathbf{k}$ vectors were generated
for a desired relative accuracy of 0.05. The Ewald parameters
were set to approximately 0.18, 0.15, and 0.11\,\AA$^{-1}$ and the number
of $\mathbf{k}$ vectors were 748, 436, and 183 for the different cutoffs.
In simulations with a PPPM solver for dispersion, we used interpolation
order $P = 5$. Following our results from Section \ref{influence},
a grid with $12\times 12\times 36$ mesh points was used
for the dispersion interactions. The Ewald parameter for dispersion was
set to $\beta$ = 0.28\,\AA$^{-1}$ for all cutoffs.
The parameters used for the long-range solver for the Coulomb interaction 
were the same as those in simulations with a plain cutoff for
dispersion.

\begin{table}[h]
\caption{Results of the validation runs for the SPC/E water. Simulation marked with a dagger
was run with higher precision.\label{SPCE}}
\scalebox{0.9}{
\begin{tabular}{l l c r r r r r}
\hline \hline
& & & & \multicolumn{3}{c}{Surface tension,
mN/m} & \\
$T$ (K)  &  solver  & $r_c$ (\AA)  &  $\rho_{\mathrm{liq}}$ (kg/L) 
& $\gamma_{p}  $  &  $\gamma_{t}  $ &  $\gamma $
& $t_c$ (ms)\\
\hline
300&cutoff       & 10.0 & 0.9882   & 54.59(100) & 5.27  & 59.86(100) &
169\\
      &                & 12.0 &  0.9918 &  56.38(100) & 3.72 &
      60.10(100) & 259 \\
      &                & 16.0 &  0.9944  & 57.51(84)  & 2.12 &
      59.63(84) & 531\\
      & Ewald     & 10.0 & 0.9964   & 59.39(100) & -       &
      59.39(100) & 455\\
     &                  & 12.0 & 0.9962  & 60.84(100) & -       &
     60.84(100) & 474 \\
     &                  & 16.0 &  0.9967 & 60.56(100) & -
     & 60.56(100)  & 756\\
     & PPPM & 10.0 & 0.9965 & 60.72(90) & -  & 60.72(90) & 229\\
     &          & 12.0 & 0.9964 &   60.11(80)   &
     -   & 60.11(80) & 364\\
     &          & 16.0  & 0.9963 & 59.64(90) & - & 59.64(90)
      &   737   \\ 
   &                  & 10.0 & 0.9964  &  61.06(80)
    & -       &   61.06(80) \dag  & -\\ 
     &                &        &               &        &         &        &                   \\
      
350 & cutoff    & 10.0 & 0.9539  & 47.40(60)  & 4.81  & 52.21(60) & 166 \\
       &              & 12.0 & 0.9576  & 48.71(60)  & 3.42  &
       52.13(60) & 254\\
       &             &  16.0 & 0.9607  & 49.78(70)  & 1.96  &
       51.74(70) & 517 \\
       & Ewald   & 10.0 & 0.9617   & 52.55(80) & -     &  52.55(80) & 448 \\
       &              & 12.0 &  0.9622   &  51.72(70) & -     &
       51.72(70) & 531 \\
       &             &  16.0 &  0.9626  &  52.13(70) & -
       & 52.13(70) & 743 \\
       & PPPM & 10.0 & 0.9629 & 53.29(60) & - & 53.29(60) & 239\\
      &           & 12.0 & 0.9631 & 52.35(70) & - & 52.35(70) & 350
      \\
      &           & 16.0 &0.9630 & 52.30(70) &  -  &52.30(70) & 737\\
      &                &        &               &            &                     &        &                   \\
      
400 & cutoff    & 10.0 &  0.9067  &  39.89(60) & 4.20  & 44.09(60) & 165 \\
       &              & 12.0 &  0.9114  &  40.42(60) & 3.02 &
       43.44(60)  & 244\\
       &             &  16.0 &  0.9151 &   41.36(58) & 1.75 &
       43.11(58) & 494 \\
       & Ewald   & 10.0 &  0.9164  &  42.97(60)  & -  & 42.97(60)  & 438 \\
       &              & 12.0 &   0.9168  &  43.71(70)  & -  &
       43.71(70) & 512 \\
       &             &  16.0 &  0.9172 & 43.34(60)   & -  &
       43.34(60)  & 716\\
       & PPPM & 10.0 & 0.9177  & 43.98(60) & - & 43.98(60) & 240\\
       &      & 12.0 & 0.9178 & 43.89(60) & - & 43.89(60) & 345\\
       &      & 16.0 &  0.9178 &  43.44(60) &     -      & 43.44(60) & 710\\
 
\hline \hline     
\end{tabular}
}
\end{table}

Table \ref{SPCE} shows the results of the simulations. 
When not using a long range solver, the simulated density shows slight
dependence on the chosen cutoff radius, whereas practically no
dependence can be observed when using a long-range solver for
dispersion. For simulated surface tensions, neither the cutoff nor the
chosen dispersion solver have a strong influence. 
The weak or non-existent influence on physical
properties of the way
dispersion interactions are calculated is due to the fact that 
Coulomb, and not dispersion, interactions are the dominant
contribution to the interactions in this system.

Again, our results are in good agreement with the majority of the literature;
\cite{intVeld.2007, Chen.2007, Wang.2009, Sakamaki.2011} 
however, they differ substantially from those reported by Shi
et al.\cite{Shi.2006}, who performed simulations of SPC/E with a PPPM 
for dispersion, too.
For example, their result for the surface tension at 300\,K 
is more than 70\,mN/m (read from Fig.\ 6 in
Ref.\ \onlinecite{Shi.2006}), whereas the surface tensions in our simulations
are always about 60\,mN/m, consistent with other studies. To ensure the validity
of our results we have run an additional simulation with increased
accuracy, in which we set the Ewald parameter for dispersion to $\beta = 0.3$~\AA$^{-1}$,
the interpolation
order to $P=5$ and the grid spacing to $h \approx 1.56$~\AA\
corresponding to $32\times 32\times 96$ mesh points. Results of
this simulation, marked with a dagger in Table \ref{SPCE}, are
in good agreement with the rest of our results. 
The increased value
for the surface tension in simulations by Shi et al. might be related to the 
small number of water molecules (800) in their simulation or the choice 
of the Ewald parameter (0.9, units not given), but is most likely
caused by their short sampling time of only 100\,000
timesteps, as substantially longer run times are required to
achieve equilibration for water at an interface.\cite{Ismail.2006}

\subsection{Hexane}
If not given in the following, all settings for the hexane simulations
were as those reported in
Section \ref{influence}.
We studied temperatures of 300, 350, and 400\,K and 
cutoffs of 10, 12, and 16\,\AA.

In simulations with a plain cutoff for dispersion, a PPPM solver was
used for electrostatics. The grid dimension was set to $20\times
20\times 45$ and the interpolation order to $P = 5$. The Ewald
parameter was approximately $\beta$ = 0.17, 0.16, and 0.14\,\AA$^{-1}$ for the
different cutoffs.
The desired precision was set to 0.05 in simulations with the Ewald
method for dispersion and Coulomb interactions. The resulting Ewald
parameters and number of $\mathbf{k}$ vectors were the same as those
in simulations with SPC/E water.
In simulations with the PPPM for dispersion, the interpolation
order was set to $P = 5$, the grid size was set to $12\times 12\times 36$
and the Ewald parameter was set to $\beta$ = 0.28~\AA$^{-1}$
in all simulations.
Coulomb interactions were treated in a same way as in simulations with
a cutoff for dispersion.

\begin{table}[h]
\caption{Results of the validation runs for the OPLS hexane\label{hexane}}
\scalebox{0.9}{
\begin{tabular}{l l c r r r r r}
\hline \hline 
& & &  & \multicolumn{3}{c}{Surface tension,
mN/m} \\
$T$ (K)  &  solver  & $r_c$ (\AA)  &  $\rho_{\mathrm{liq}}$ (kg/L)  
 & $\gamma_{p}  $  &  $\gamma_{t}  $ &  $\gamma $ & $t_c$ (ms)
\\
\hline
300 & cutoff    & 10.0 & 0.6058  & 7.83(50)  & 4.59 & 12.42(50) & 137  \\
       &              & 12.0 & 0.6251 & 10.21(50) & 3.75 & 13.96(50) &
       207\\
       &              &  16.0 & 0.6367  & 13.00(50) & 2.34 & 15.34(50)
       & 421 \\
       & Ewald   & 10.0 & 0.6368    & 14.40(50)  & - & 14.40(50) & 409 \\
       &              & 12.0 &  0.6385  & 14.91(50)  &  - & 14.91(50)
       & 414 \\
       &             &  16.0 &  0.6410   & 14.91(56)  & -  & 14.91(56)
       & 634\\
      & PPPM& 10.0 &  0.6434 & 16.41(50) &  -  & 16.41(50) & 201\\
       &         & 12.0 & 0.6439  & 16.16(50) &  -  & 16.16(50) & 352 \\
       &           &  16.0 & 0.6453 & 15.89(50) & - & 15.89(50) & 601\\
      &                &        &               &            &                    &        &                   \\
      
350 & cutoff    & 10.0 &  0.5237  & 2.18(40)  & 2.13  & 4.31(40) & 118  \\
       &              & 12.0 &  0.5534  & 4.78(40)  & 2.20  & 6.98(40)
       & 182\\
       &             &  16.0 &  0.5721  & 7.44(50)  & 1.71 & 9.15(50)
       & 383 \\
       & Ewald   & 10.0 & 0.5722  & 8.09(50)  & - & 8.09(50) & 384 \\
       &              & 12.0 & 0.5778    & 8.52(40)  & - &  8.52(40) &
       438\\
       &             &  16.0 &  0.5805  & 9.03(40)  & - & 9.03(40) & 575 \\
      &  PPPM & 10.0 &  0.5823  & 9.97(44) & -    & 9.97(44) & 183 \\
       &              & 12.0 & 0.5839 & 9.77(60) & - & 9.77(60)  & 276\\
       &             &  16.0 & 0.5851 & 9.89(44) & - & 9.89(44) & 547\\
      &                &        &               &             &                    &        &                   \\
      
400 & cutoff    & 10.0 &  n.a  &  -1.55(32) & n.a. & -1.55(32) & 92 \\
       &              & 12.0 &  0.4467  &  0.30(36) & 0.74 & 1.04(36)
       & 149 \\
       &             &  16.0 &  0.4905  &  1.83(36) & 0.85 & 2.68(36)
       & 316 \\
       & Ewald   & 10.0 & 0.4881    & 2.26(50)  & - & 2.26(50) & 368 \\
       &              & 12.0 &  0.5037  &3.07(32)  & - & 3.07(16)  & 402\\
       &             &  16.0 &  0.5039  & 3.43(40)  & - & 3.43(40) & 498\\
      &    PPPM  & 10.0 &  0.5099  & 4.59(36)  &  -& 4.59(36) & 158\\
       &              & 12.0 & 0.5106 & 4.66(40)    &  - & 4.66(40)  &
       234\\
       &             &  16.0 & 0.5157 & 4.45(46) & - & 4.45(46) & 452
       \\
\hline \hline
\end{tabular}
}
\end{table}

\begin{figure}
\includegraphics[scale=1]{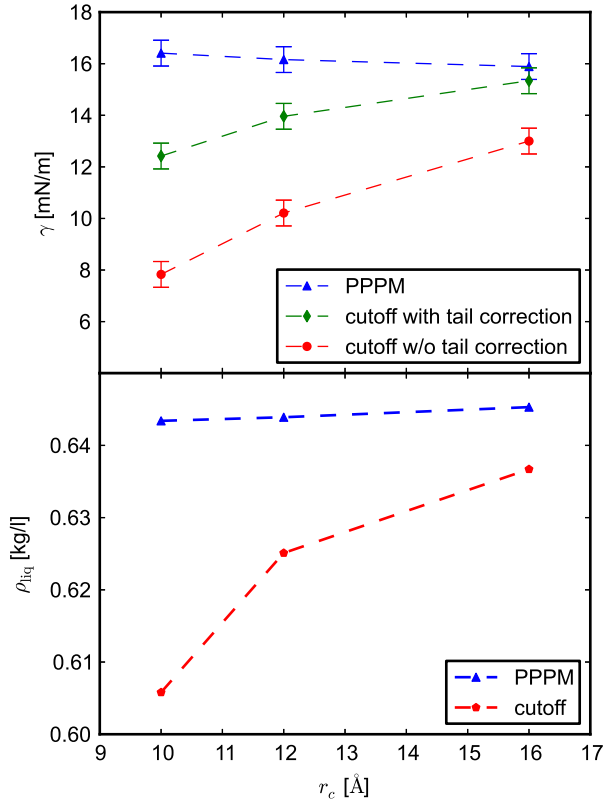}
\caption{Surface tensions and densities simulated when using a PPPM
  or a plain cutoff for dispersion. Results obtained when not using a
  long-range solver strongly depend on the chosen cutoff and approach
  the results of the simulation with the PPPM with increasing cutoff size.
\label{hexane_cutoff_dependence}}
\end{figure}

The results are summarized in Table \ref{hexane}. 
The chosen cutoff radius has a strong influence on the results in simulations
without a long-range dispersion solver. In contrast, the results for Ewald
summation show weak and the results for the PPPM show no dependence on the
chosen cutoff radius.
Our results for the PPPM are in good agreement with those from Ismail et al.\cite{Ismail.2007}
in simulations with an Ewald sum for dispersion. This, and the fact that the chosen cutoff
does not influence the results, confirms the validity of our simulations and
the good choice of the Ewald and grid parameters.
Our simulations with Ewald sums
provide lower surface tensions, which is caused by insufficient accuracy 
in these simulations. As can be seen from Figure
\ref{hexane_cutoff_dependence}, the simulation results when not using
a long-range dispersion solver approach those obtained with the PPPM
when increasing the cutoff. However, even those with 
a cutoff of 16\,\AA\ provide surface tensions and densities that are below those
obtained from PPPM simulations.

\section{Performance Comparison}
\label{performance}

To measure the simulation time, each of the simulations in Section
\ref{application} was run for 1\,000 timesteps on a single core. The
resulting computation times per timestep $t_c$ are given in the last
column of Tables \ref{LJ} to \ref{hexane}. The simulations were
executed on Intel Harpertown E5454 processors with eight 3.0 GHz Xeon cores.

For a fair comparison between the different solvers, one should
consider different solvers at the same temperature with the cutoff
that provides the results that are obtained in the limit of high
accuracy simulations. If for a given solver and temperatures different
cutoffs provided accurate results, the fastest of those simulations
should be used.

A quick comparison of the PPPM with the Ewald shows that simulations
with the PPPM were faster in all cases. 
As the Ewald sum was always slower than the PPPM, we omit comparisons
between the Ewald solver and the plain cutoff and continue with
comparing simulations with a PPPM to those with a plain cutoff. 
For the LJ system, accurate surface tensions and densities were
obtained only in simulation in which the cutoff
was $r_c = 7.5\sigma$ in simulations without long-range dispersion solvers. 
For simulations with the PPPM for dispersion $r_c = 3\sigma$ provides
proper results with maximum efficiency. Comparing the corresponding
simulation times shows the computational superiority of using the
PPPM for dispersion.

In simulations with water, the results of the comparison are different
when examining the simulated surface tensions or simulated densities.
As surface tensions were approximately the same throughout all
simulations, the
 PPPM was outperformed by the plain cutoff for this comparison.
The reason for the observed behavior is not the choice of parameters, but the
dominance of Coulomb interaction in the examined system. When comparing the 
simulation times it has to be kept in mind that the simulations with a plain
cutoff are incorrect during the simulations and are only corrected {\em a
posteriori}.
If this correction is not possible after the simulations, or if a correct value
of the surface tension is required during the simulations for any reason,
a larger cutoff is required in simulations with a plain cutoff.
Simulations with a PPPM are more efficient in such cases. The increase
in simulation time is about 35 percent when using the PPPM for
dispersion compared to a plain cutoff. 

If highly accurate simulated densities are important, then the cutoff for dispersion
interactions should be chosen to be at least $r_c = 12$\,\AA\ in
simulations without long-range dispersion solvers. 
Comparing the simulation time of these simulations to those with a PPPM with a
cutoff $r_c = 10$\,\AA\ shows that using the PPPM is computationally
favorable in this case. 

For simulations of hexane, in which dispersion interactions dominate,
the largest cutoff has to be selected in simulations without long-range 
dispersion solvers, whereas a small cutoff is sufficient in simulations with
the PPPM. As a consequence, the simulations with the PPPM were much faster than those
without a long-range dispersion solver.

We would like to note that simulations with the long-range dispersion
solver were run without tabulating the pair potential. Including this
tabulation will results in additional speed-up of the simulations and
might change the comparison above.

\section{Conclusions and Outlook}
\label{conclusions}
We present a PPPM algorithm for dispersion interactions that calculates
long-range dispersion forces accurately and efficiently for inhomogeneous
systems. When used correctly, this solver strongly reduces the error
that is caused when truncating the pair potential at a plain cutoff
and thus provides a better description of the physics of a simulated
system.

We derived and tested error estimates that describe the effect of the
parameters of the PPPM on simulation results. The presented error
estimates are only valid for bulk phase systems and should not be
relied on when simulated energies or pressures are relevant.

For not having to rely on the presented error measures, we explored
the parameter space to provide parameters that can be used in future
simulations. For
real physical systems of surfaces, a combination of the Ewald parameter $\beta =
0.28$\,\AA$^{-1}$,  the interpolation order $P = 5$, the grid
spacing $h \approx 4.17$\,\AA, and the real space cutoff $r_c =
10.0$\,\AA\ provided good results for different materials at
different temperatures.

We have applied the developed algorithm to study
the surface tension of LJ particles, SPC/E water,
and hexane. 
The described algorithm outperforms 
Ewald summation for all systems that were examined
in this study in terms of simulation time. 

Comparing the PPPM to simulations with a plain cutoff show that the
PPPM is favorable when correction methods cannot be applied or
correction methods do not work properly, as for example near the
critical point or in some of our hexane and LJ simulations. 
In systems that are dominated by dispersion, the PPPM outperforms
simulations without long-range solvers in tems of computation time, as 
latter simulations require larger cutoffs. 

For strongly charged systems, the PPPM provides a benefit in
simulation time only if densities are needed at a high
accuracy. In other cases a plain cutoff is favorable in terms of
computation time here. However, the advantage of correctness during the simulation
and the applicability to arbitrarily shaped surfaces remains.

\begin{acknowledgements}
We would like to thank
Stan Moore (Brigham Young) and Paul Crozier and Steve Plimpton from
Sandia National Laboratories for their support.  

Financial support from the Deutsche Forschungsgemeinschaft (German Research Association) through Grant GSC 111 is gratefully acknowledged.
\end{acknowledgements}

\begin{appendix}
\end{appendix}

%
%

%


\bibliographystyle{aipnum4-1}
%

\end{document}